\documentclass[12pt]{article}
\usepackage{amsfonts}
\usepackage{amsmath}
\usepackage{amssymb}
\usepackage{graphicx}
\usepackage{xcolor}
\usepackage[all, knot]{xy}
\usepackage{tikz}
\usepackage{cancel}
\usepackage{comment}

\usepackage{ulem}
\usepackage{hyperref}

\usepackage[utf8]{inputenc}
\usepackage{epstopdf}
\usepackage[footnotesize]{caption}
\usepackage{amsthm}
\usepackage{enumitem}
\usepackage{mathrsfs}

\makeatletter
\newcommand\xleftrightarrow[2][]{%
	\ext@arrow 9999{\longleftrightarrowfill@}{#1}{#2}}
\newcommand\longleftrightarrowfill@{%
	\arrowfill@\leftarrow\relbar\rightarrow}
\makeatother

\usepackage[margin=3cm]{geometry}

\def \be {\begin{equation}}
\def \ee {\end{equation}}
\def \bea {\begin{eqnarray}}
\def \eea {\end{eqnarray}}
\def \nn {\nonumber}

\def \rr {\raise.35ex\hbox{\small $\prime$}\kern-.17em{\mbox{\large $\imath$}}}

\def \dels {\partial\kern-.6em /\kern.1em}
\def \As {{A\kern-.5em / \kern.5em}}
\def \Ds {D\kern-.7em / \kern.5em}

\def \ks {k\kern-.5em /}
\def \ls {l\kern-.5em /}

\def \sgn {\mbox{\small sgn}}

\newcommand{\er}[1]{\eqref{#1}}

\newcommand{\ci}[1]{}

\newcommand{\ba}{\begin{eqnarray}}
\newcommand{\ea}{\end{eqnarray}}
\newcommand{\bal}{\begin{align}}
\newcommand{\eal}{\end{align}}
\newcommand{\bay}[1]{\left(\begin{array}{#1}}
\newcommand{\eay}{\end{array}\right)}

\def\xD{{\Delta}}

\newcommand{\hide}[1]{}

\newlist{axioms}{enumerate}{2}
\setlist[axioms,1]{label=\textbf{A\arabic{axiomsi}.}, ref=A\arabic{axiomsi}}
\setlist[axioms,2]{label=\textbf{A\arabic{axiomsi}\rlap{\myEnumCounter{axiomsii}}.},%
                   ref=A\arabic{axiomsi}\myEnumCounter{axiomsii},%
                   align=parleft,%
                   leftmargin=0em,%
                   itemsep=1.4ex,%
                   before={\stepcounter{axiomsi}}}

  \usetikzlibrary{decorations.markings}

\begin{document}

\begin{titlepage}
\begin{center}

\textbf{\LARGE
Higher-Dimensional Fermionic SYK Model in IR Region
\vskip.3cm
}
\vskip .5in
{\large
Xing Huang$^{a, b}$ \footnote{e-mail address: xingavatar@gmail.com} and
Chen-Te Ma$^{c}$ \footnote{e-mail address: yefgst@gmail.com}, 
\\
\vskip 1mm
}
{\sl
$^a$
Institute of Modern Physics, Northwest University, Xi'an 710069, China.
\\
$^b$
NSFC-SPTP Peng Huanwu Center for Fundamental Theory, Xi'an 710127, China.
\\
$^c$ 
Department of Physics, Great Bay University, Dongguan, Guangdong 52300, China
}\\
\vskip 1mm
\vspace{40pt}
\end{center}

\begin{abstract} 
\noindent
We study the 2D fermionic SYK model with Majorana fermions, featuring a quartic kinetic term and a $2q$-body interaction with Gaussian disorder. 
By minimizing the effective action or solving the SD equation for $q=1$, we determine that the appropriate ansatz involves zero spins.
Our computation of the Lyapunov exponent shows violations of chaos and unitarity bounds.
The gravitational dual corresponds to AdS$_3$ Einstein gravity with a finite radial cut-off, even if we lose the non-zero spins.
We also extend the SYK model to higher dimensions while maintaining a similar SD equation in the IR.
\end{abstract}
\end{titlepage}

\section{Introduction}
\label{sec:1}
\noindent
The Anti-de Sitter/Conformal Field Theory (AdS/CFT) correspondence is a conjecture about the duality of the physical observables between two theories \cite{Maldacena:1997re}.
One side is a $(D+1)$-dimensional gravitational theory in AdS space.
The dual CFT lives on the $D$-dimensional boundary of the AdS$_{D+1}$ space.
A leading candidate for a consistent theory of quantum gravity compatible with the AdS/CFT correspondence is String Theory \cite{Gubser:1998bc}.
The AdS/CFT dictionary provides a crucial framework for understanding the emergence of spacetime from CFT data in the semi-classical regime \cite{Witten:1998qj}.
\\

\noindent
The challenges of ultraviolet (UV) completion in gravitational theories stem from the dimensionful nature of the gravitational constant.
It highlights lower-dimensional cases, such as 3D pure Einstein gravity and 2D pure Jackiw-Teitelboim (JT) gravity \cite{Almheiri:2014cka}, where simplifications occur.
These models have a boundary description \cite{Freidel:2008sh,Jensen:2016pah,Maldacena:2016upp,Engelsoy:2016xyb}, which imposes constraints on the boundary theory.
The discussion then connects to the AdS/CFT correspondence, where a full Lagrangian description of gravity and the boundary theory is difficult to obtain.
Instead, focusing on the UV-complete boundary theory provides a well-controlled semi-classical limit, in which graviton and matter fields emerge in different orders, aligning with the emergent spacetime.
\\

\noindent
The Sachdev-Ye-Kitaev (SYK) model with $N$ Majorana fermions \cite{Polchinski:2016xgd}
\begin{eqnarray}
H_{\mathrm{SYK}} = -i^{\frac{q}{2}} \sum_{1 \le i_1 < i_2 < \cdots < i_q \le N} J_{i_1 i_2 \cdots i_q} \psi_{i_1} \psi_{i_2} \cdots \psi_{i_q},
\end{eqnarray}
where $\psi_j$ are the Majorana fermions that satisfy the anticommutation relations
\bea
\{\psi_j, \psi_k\}=\delta_{jk},
\eea
and $J_{i_1 i_2 \cdots i_q}$ are random coupling constants that follow a Gaussian distribution with zero mean and variance, showing the emergence of AdS$_2$ JT pure gravity and the matter fields in the large-$N$ expansion at the infrared (IR) regime \cite{Jevicki:2016bwu,Maldacena:2016hyu}.
The effective action in the IR regime is described by the Schwarzian derivative, where
\bea
\{f, \tau\}\equiv\frac{f^{\prime\prime\prime}(\tau)}{f^{\prime}(\tau)}-\frac{3}{2}\frac{f^{\prime\prime 2}(\tau)}{f^{\prime 2}(\tau)},
\eea
which is associated with broken conformal symmetry \cite{Maldacena:2016hyu}.
The SYK model features an emergent conformal symmetry.
Holographic analysis is often restricted to CFTs \cite{Dolan:2000ut,Rychkov:2009ij}, although it could be extended to encompass UV-complete quantum field theories (QFTs).
This restriction is unnecessary when exploring the nearly AdS/CFT correspondence.
The computation of the out-of-time ordered correlator (OTOC) shows the exponential growth, and the exponent saturates the chaos bound \cite{Polchinski:2016xgd,Shenker:2013pqa,Maldacena:2015waa}.
The SYK model is a fascinating playground for exploring quantum chaos and holography.
\\

\noindent
The motivation for 2D generalization stems from the nearly AdS/CFT correspondence in 3D Einstein gravity, which necessitates a 2D generalization of the SYK model.
The Lorentzian action of the 2D fermionic SYK model consists of a kinetic term $S_{\mathrm{UV}}$ and an interaction term $S_{\mathrm{IR}}$ \cite{Turiaci:2017zwd}
\bea
S_{\mathrm{2D}}=S_{\mathrm{UV}}+S_{\mathrm{IR}},
\eea
where
\bea
S_{\mathrm{UV}}&=&\sum_{j=1}^N\int d^2x\ \epsilon^{\mu\nu}\psi_+^j(\partial_{\mu}\psi_+^j)\psi_-^j(\partial_{\nu}\psi_-^j);
\nn\\
S_{\mathrm{IR}}&=&i^q\sum_{1\le i_<  i_2<\cdots<i_q<j_1<j_2<\cdots<j_q\le N}\int d^2x \ J_{i_1,i_2, \cdots, i_q, j_1, j_2, \cdots, j_q}
\nn\\
&&\times
\psi_-^{i_1}\psi_-^{i_2}\cdots \psi_-^{i_q}\psi_+^{j_1}\psi_+^{j_2}\cdots\psi_+^{j_q}.
\nn\\
\eea
$\psi_+$ and $\psi_-$ are the two chiral components of a 2D Majorana fermion.
$J_{i_1,i_2, \cdots, i_q, j_1, j_2, \cdots, j_q}$ follows a Gaussian random distribution with
\bea
\langle J_{i_1,i_2, \cdots, i_q, j_1, j_2, \cdots, j_q}J_{i_1,i_2, \cdots, i_q, j_1, j_2, \cdots, j_q}\rangle=i\frac{J^2(q-1)!q!}{N^{2q-1}},
\eea
where $J$ is a fixed coupling constant.
\\

\noindent
The kinematic term $S_{\mathrm{UV}}$ includes two derivatives, ensuring that the fermion fields remain dimensionless.
This leads to a fixed coupling constant with a positive energy dimension of $[J]=2$, which ultimately supports renormalizability.
The action preserves local Lorentz symmetry and area-preserving diffeomorphisms.
At the same time, the interaction term dominates in the IR limit $N\gg\beta^2 J\gg1$ \cite{Turiaci:2017zwd}, where $\beta$ is the inverse temperature.
The model exhibits reparametrization symmetry in the IR, characterized by a transformation law for the fermionic fields \cite{Turiaci:2017zwd}
\bea
\psi_{\pm}^j(x)\rightarrow \bigg|\det\frac{\partial y^{\mu}}{\partial x^{\nu}}\bigg|^{\frac{1}{2q}}\psi_{\pm}^j(y).
\eea
The UV limit is topological {\it without} the UV divergence \cite{Turiaci:2017zwd}.
This generalization does not require that the spatial dimension be discrete, unlike other generalizations \cite{Gu:2016oyy,Berkooz:2016cvq}.
A crucial feature of this model is that the large-$N$ limit leads to a Schwinger-Dyson (SD) equation with an undetermined spin parameter $s$ \cite{Turiaci:2017zwd}.
However, it needs to be determined, as the two-point function is unambiguous.
The IR effective action, expressed in terms of double Schwarzian derivatives \cite{Turiaci:2017zwd}, suggests a connection to AdS$_3$ Einstein gravity with a finite radial cutoff \cite{Freidel:2008sh}.
This is indicative of a $T\bar{T}$ deformation \cite{Smirnov:2016lqw,Cavaglia:2016oda,McGough:2016lol}.
This intriguing result depends on the {\it specific} value of $s$.
Without confusing other generalizations \cite{Murugan:2017eto}, we call this generalization {\it fermionic} SYK model because the field content {\it only} contains the Majorana fermion fields.
The central question we aim to address in this letter is: {\it What are the IR properties of the two-dimensional fermionic SYK model, and how does it extend to higher dimensions?}
\\

\noindent
In this letter, we explore the IR solutions of the large-$N$ SD equation in the 2D fermionic SYK model and extend the analysis to higher dimensions.
Key findings include the exact solution for $q=1$, which shows that there are no spins in the 2D case.
Minimizing the effective action further confirms this.
The Lyapunov exponent $\lambda_{\mathrm{L}}$ for the OTOC
\bea
{\mathcal F}_{ab}(x_1, x_2, x_3, x_4)=\frac{1}{N^2}\sum_{j, k=1}^N
\langle \psi_a^j(x_1)\psi_a^j(x_2)\psi_b^k(x_3)\psi_b^k(x_4)\rangle\sim e^{\lambda_\mathrm{L} t}
\eea
is computed, revealing a violation of the chaos bound.
This violation is attributed to the breakdown of unitarity at IR.
We analyze the unitary bound using the bootstrap, and it has been applied to the SYK-like model to examine unitarity \cite{Chang:2021fmd,Chang:2021wbx}.
These apparent violations may arise from neglecting renormalization effects in the IR.
Our computation does not justify the non-unitarity observed in the ultraviolet (UV) region.
However, restoring the IR unitarity may require the supersymmetry \cite{Murugan:2017eto,Peng:2018zap,Chang:2021fmd,Chang:2021wbx,Chang:2023gow}, and the unitarity issue in the UV regime may not arise.
Furthermore, the effective action in the interacting representation corresponds to AdS$_3$ Einstein gravity with a finite cutoff, even when considering a zero-spin variable.
Finally, our study extends the fermionic SYK model to higher dimensions, showing that the self-consistency equation can be solved in the same manner as in lower dimensions.

\section{2D Fermionic SYK Model}
\label{sec:2}
\noindent
The partition function is
\bea
&&
Z_{\mathrm{2D}}
\nn\\
&=&\int d J_{i_1, i_2, \cdots, i_q, j_1, j_2, \cdots, j_q}\int D\psi_+D\psi_-\ \exp(iS_{\mathrm{2D}})
\nn\\
&&\times
\exp\bigg(-\sum_{1\le i_<  i_2<\cdots<i_q<j_1<j_2<\cdots<j_q\le N} J_{i_1, i_2, \cdots, i_q, j_1, j_2, \cdots, j_q}^2
\frac{-iN^{2q-1}}{2J^2(q-1)!q!}\bigg).
\nn\\
\eea
Integrating the random coupling constant is equivalent to using the following equation of motion
\bea
J^{i_1, i_2, \cdots, i_q, j_1, j_2, j_q}=-i^q\frac{J^2(q-1)!q!}{N^{2q-1}}\int d^2x\
\psi_-^{i_1}\psi_-^{i_2}\cdots\psi_-^{i_q}\psi_+^{j_1}\psi_+^{j_2}\cdots\psi_+^{j_q}.
\nn\\
\eea
The partition function can be expressed in the following way:
\bea
Z_{\mathrm{2D}}&=&\int D\psi_+D\psi_-\ \exp\bigg\lbrack iS_{\mathrm{UV}}
\nn\\
&&
-i\frac{J^2N}{2q}\int d^2xd^2\tilde{x}\ \bigg(
\frac{1}{N}\sum_{l_1=1}^N\psi_+^{l_1}(x)\psi_+^{l_1}(\tilde{x})
\bigg)^q
\bigg(
\frac{1}{N}\sum_{l_2=1}^N\psi_-^{l_2}(x)\psi_-^{l_2}(\tilde{x})
\bigg)^q
\bigg\rbrack.
\nn\\
\eea
\\

\noindent
Now we insert the identity:
\bea
&&
1
\nn\\
&=&\int D\tilde{G}_+D\tilde{G}_-\
\delta\bigg(\tilde{G}_+(x, \tilde{x})-\frac{1}{N}\sum_{l_1=1}^N\psi_+^{l_1}(x)\psi_+^{l_1}(\tilde{x})\bigg)
\nn\\
&&\times
\delta\bigg(\tilde{G}_-(x, \tilde{x})-\frac{1}{N}\sum_{l_2=1}^N\psi_-^{l_2}(x)\psi_-^{l_2}(\tilde{x})\bigg)
\nn\\
&\sim&\int D\tilde{G}_+D\tilde{G}_-D\tilde{\Sigma}_+D\tilde{\Sigma}_-\
\exp\bigg\lbrack i\int d^2xd^2\tilde{x}\ \frac{\tilde{\Sigma}_+(x, \tilde{x})}{2}
\nn\\
&&\times
\bigg(N\tilde{G}_+(x, \tilde{x})-\sum_{l_1=1}^N\psi_+^{l_1}(x)\psi_+^{l_1}(\tilde{x})\bigg)
\nn\\
&&
+i\int d^2xd^2\tilde{x}\ \frac{\tilde{\Sigma}_-(x, \tilde{x})}{2}
\bigg(N\tilde{G}_-(x, \tilde{x})-\sum_{l_1=1}^N\psi_-^{l_1}(x)\psi_-^{l_1}(\tilde{x})\bigg)
\bigg\rbrack
\eea
to the partition function
\bea
Z_{\mathrm{2D}}&=&\int D\psi_+D\psi_-D\tilde{G}_+D\tilde{G}_-D\tilde{\Sigma}_+D\tilde{\Sigma}_-\ \exp\bigg\lbrack iS_{\mathrm{UV}}
\nn\\
&&
-i\frac{J^2N}{2q}\int d^2xd^2\tilde{x}\ \bigg(
\frac{1}{N}\sum_{l_1=1}^N\psi_+^{l_1}(x)\psi_+^{l_1}(\tilde{x})
\bigg)^q
\bigg(
\frac{1}{N}\sum_{l_2=1}^N\psi_-^{l_2}(x)\psi_-^{l_2}(\tilde{x})
\bigg)^q
\bigg\rbrack
\nn\\
&&\times
\exp\bigg\lbrack i\int d^2xd^2\tilde{x}\ \frac{\tilde{\Sigma}_+(x, \tilde{x})}{2}
\bigg(N\tilde{G}_+(x, \tilde{x})-\sum_{l_1=1}^N\psi_+^{l_1}(x)\psi_+^{l_1}(\tilde{x})\bigg)
\nn\\
&&
+i\int d^2xd^2\tilde{x}\ \frac{\tilde{\Sigma}_-(x, \tilde{x})}{2}
\bigg(N\tilde{G}_-(x, \tilde{x})-\sum_{l_1=1}^N\psi_-^{l_1}(x)\psi_-^{l_1}(\tilde{x})\bigg)
\bigg\rbrack.
\eea
We introduce $e_{\mu}^{\pm}$ to have the quadratic fermion fields
\bea
Z_{\mathrm{2D}}&=&\int D\psi_+D\psi_-D\tilde{G}_+D\tilde{G}_-D\tilde{\Sigma}_+D\tilde{\Sigma}_-De_{\mu}^+De_{\nu}^-\ \exp\bigg\lbrack iS_{\mathrm{UV1}}
\nn\\
&&
-i\frac{J^2N}{2q}\int d^2xd^2\tilde{x}\ \bigg(
\frac{1}{N}\sum_{l_1=1}^N\psi_+^{l_1}(x)\psi_+^{l_1}(\tilde{x})
\bigg)^q
\bigg(
\frac{1}{N}\sum_{l_2=1}^N\psi_-^{l_2}(x)\psi_-^{l_2}(\tilde{x})
\bigg)^q
\bigg\rbrack
\nn\\
&&\times
\exp\bigg\lbrack i\int d^2xd^2\tilde{x}\ \frac{\tilde{\Sigma}_+(x, \tilde{x})}{2}
\bigg(N\tilde{G}_+(x, \tilde{x})-\sum_{l_1=1}^N\psi_+^{l_1}(x)\psi_+^{l_1}(\tilde{x})\bigg)
\nn\\
&&
+i\int d^2xd^2\tilde{x}\ \frac{\tilde{\Sigma}_-(x, \tilde{x})}{2}
\bigg(N\tilde{G}_-(x, \tilde{x})-\sum_{l_1=1}^N\psi_-^{l_1}(x)\psi_-^{l_1}(\tilde{x})\bigg)
\bigg\rbrack,
\eea
where
\bea
S_{\mathrm{UV1}}\equiv\frac{1}{2}\int d^2x\ \epsilon^{\mu\nu}\bigg(\sum_{a=\pm}\sum_{j=1}^Ne_{\mu}^a\psi_a^j\partial_{\nu}\psi_a^j-\sum_{a, b=\pm}\epsilon_{ab}e_{\mu}^ae_{\nu}^b\bigg).
\eea
Because we only have the quadratic fermion fields, we can perform the Gaussian integration to integrate the fermion fields
\bea
&&
Z_{\mathrm{2D}}
\nn\\
&=&\int D\tilde{G}_+D\tilde{G}_-D\tilde{\Sigma}_+D\tilde{\Sigma}_-De_{\mu}^+De_{\nu}^-\
\exp\bigg\{iN\bigg\lbrack\sum_{a=\pm}\bigg(-\ln\mathrm{Pf}(\epsilon^{\mu\nu} e_{\mu}^a\partial_{\nu}-\tilde{\Sigma}^a\big)
\nn\\
&&
+\frac{1}{2}\int d^2xd^2\tilde{x}\ \tilde{\Sigma}^a\tilde{G}_a\bigg)
-\epsilon^{\mu\nu}\int d^2x\ e_{\mu}^+e_{\nu}^-
-\frac{J^2}{2q}\int d^2xd^2\tilde{x}\ (\tilde{G}_+)^q(\tilde{G}_-)^q
\bigg\rbrack
\bigg\}.
\nn\\
\eea
Therefore, we derive an effective action
\bea
\frac{S_{\mathrm{eff}}}{N}&=&
-\frac{1}{2}\sum_{a=\pm}\bigg(\ln\det(\epsilon^{\mu\nu} e_{\mu}^a\partial_{\nu}-\tilde{\Sigma}^a\big)\bigg)
-\epsilon^{\mu\nu}\int d^2x\ e_{\mu}^+e_{\nu}^-
\nn\\
&&
+\frac{1}{2}\sum_{a=\pm}\bigg(\int d^2xd^2\tilde{x}\ \tilde{\Sigma}^a\tilde{G}_a\bigg)
-\frac{J^2}{2q}\int d^2xd^2\tilde{x}\ (\tilde{G}_+)^q(\tilde{G}_-)^q.
\eea
We can derive the large-$N$ SD equation from the effective action and discuss the solution and the kernel.

\subsection{Large-$N$ SD Equation}
\noindent
The variation of $\tilde{\Sigma}^a$ gives the equation of motion in momentum space
\bea
\frac{1}{\bar{G}_a(k)}=-i\epsilon^{\mu\nu}e_{\mu}^a k_{\nu}-\bar{\Sigma}_a(k),
\label{SD1}
\eea
where
\bea
\bar{G}(k)\equiv\int d^2x\ e^{ik(x-x^{\prime})}\tilde{G}(x, x^{\prime}).
\eea
The variation of $\tilde{G}^{\pm}$ gives the equation of motion
\bea
\tilde{\Sigma}^{\pm}(x)=J^2\tilde{G}_{\pm}^{q-1}(x)\tilde{G}_{\mp}^q(x).
\label{SD2}
\eea
When we are only concerned with the IR limit, we can drop the derivative term to obtain
$\bar{G}_{\pm}\bar{\Sigma}_{\pm}\approx -1$.
\\

\noindent
We consider the conformal solutions as the ansatz \cite{Turiaci:2017zwd}:
\bea
\tilde{G}_{\pm}=b\frac{\sgn(x^{\pm})}{|x^+|^{\Delta\pm s}|x^-|^{\Delta\mp s}}; \qquad
\tilde{\Sigma}_{\pm}=J^2b^{2q-1}\frac{\sgn(x^{\pm})}{|x^+|^{2-\Delta\mp s}|x^-|^{2-\Delta\pm s}},
\nn\\
\eea
where $b$ is a constant and $x^{\pm}\equiv t\pm x$.
$\Delta$ denotes the conformal dimension.
To simplify the notation, we introduce $\Delta_{\pm}\equiv\Delta\pm s$.
The sign function is defined as follows
\bea
\sgn(x)=\left\{\begin{array}{ll}
1, & \mbox{if $x> 0$}. \\
0, & \mbox{if $x=0$}.\\
-1, & \mbox{if $x<0$}.
\end{array} \right.
\eea
When substituting the ansatz to Eq. \eqref{SD2}, we determine the conformal dimension $\Delta=1/q$ \cite{Turiaci:2017zwd}.
For another equation of motion, we determine the coefficient $b$ up to the derivative corrections \cite{Turiaci:2017zwd}
\bea
J^2b^{2q}\approx \frac{(1-\Delta_+)(1-\Delta_-)}{4\pi^2\tan\big(\frac{\pi}{2}\Delta_{+}\big)\cot\big(\frac{\pi}{2}\Delta_{-}\big)}.
\eea
Thus, the spin $s$ remains undetermined \cite{Turiaci:2017zwd}, but it should not have ambiguity once we find the exact solution.
\\

\noindent
We solve the SD equation in momentum space for $q=1$:
\bea
\bar{\Sigma}^{\pm}&=&J^2\bar{G}_{\mp};
\nn\\
1&=&-2i(e_+^{\pm}k_--e_-^{\pm}k_+)\bar{G}_{\pm}-\bar{\Sigma}_{\pm}\bar{G}_{\pm},
\eea
in which we use the notation $\epsilon^{+-}=2$.
The exact solution for $\bar{G}_{\pm}$ is given by:
\bea
\bar{G}_{+}=\frac{A_{+}\pm\sqrt{A_{+}^2-4J^2\frac{A_{+}}{A_-}}}{2J^2\frac{A_{+}}{A_{-}}}; \
\bar{G}_{-}=\frac{A_{+}\pm\sqrt{A_{+}^2-4J^2\frac{A_{+}}{A_-}}}{2J^2},
\eea
where $A_{\pm}\equiv -2i(e_+^{\pm}k_--e_-^{\pm}k_+)$.
When concerning the IR limit or $A_{\pm}\rightarrow 0$, we obtain that $\bar{G}_{+-}$ is independent of $k_{\pm}$.
Our ansatz in the momentum space is proportional to $k_{\pm}^{\Delta_{\pm}-1}$.
Combining two results implies that
\bea
\Delta_{\pm}=\Delta=1=\frac{1}{q}; \ s=0.
\eea
\\

\begin{figure}
\centering
\includegraphics[width=1.\linewidth]{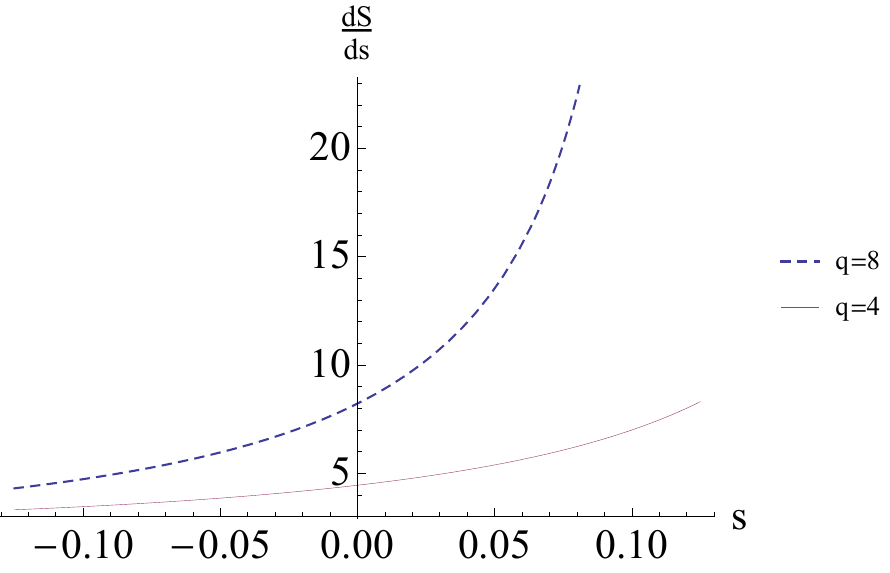}
\caption{The derivative of $S\equiv\ln(\bar{G}_+\bar{G}_-)$ for the spin $s$.
The $S$ is proportional to the effective action.
As we can see, it is neither even nor odd concerning $s$.}
\label{ssym}
\end{figure}

\noindent
For a general $q$, we can also determine the spin variable $s$ by minimizing the effective action for the spin.
The key observation is that the ansatz in Ref. \cite{Turiaci:2017zwd} makes the effective action asymmetric with respect to the spin.
In Fig. \ref{ssym}, we show that the derivative of the effective action for the spin is not symmetric.
However, we have another solution with a different ansatz
\bea
\tilde{G}_{\pm, A}=b\frac{\sgn(x^{\mp})}{|x^+|^{\Delta_{\pm}}|x^-|^{\Delta_{\mp}}}\,.
\eea
Such an extra solution (different from the previous one by $s \to - s$) essentially contributes another saddle point to the partition function. Therefore, the sum of two saddles restores the symmetry in $s$ and guarantees that the most dominant contribution to the effective action happens at $s=0$ for a general $q$.
Moreover, the choice of $s=0$ results in a more suitable ansatz in the IR limit, and $\tilde{G}_{\pm}$ and $\tilde{G}_{\pm, A}$ become the same.

\subsection{Kernel}
\noindent
The variation of Eq. \eqref{SD1} gives
\bea
\int d^2y\ \big(\delta\tilde{G}_{\pm}(x, y)\tilde{\Sigma}_{\pm}(y, z)+\tilde{G}_{\pm}(x, y)\delta\tilde{\Sigma}_{\pm}(y, z)\big)=0.
\eea
Hence the kernel, $K_{ab}$, is given by
\bea
\delta\tilde{G}_a(x_1, x_2)=\sum_{b=\pm}\int d^2x_3d^2x_4\ K_{ab}(x_1, x_2, x_3, x_4)\delta\tilde{G}_b(x_3, x_4),
\eea
where
\bea
K_{ab}&=&-J^2(q-\delta_{ab})\tilde{G}_a(x_1, x_3)\tilde{G}_a(x_2, x_4)L_{ab}(x_3, x_4);
\nn\\
L_{ab}(x, y)&=&\frac{\tilde{G}_+^q(x, y)\tilde{G}_-^q(x, y)}{\tilde{G}_a(x, y)\tilde{G}_b(x, y)}.
\eea
The kernel acting on the eigenfunction takes the value \cite{Turiaci:2017zwd}
\bea
\label{kernel}
\frac{1}{\alpha_{ab}}
\begin{pmatrix}
k_{\Delta_+}(h)\tilde{k}_{\Delta_-}(\bar{h})&\frac{q}{q-1}\tilde{k}_{\Delta_-}(h)k_{\Delta_+}(\bar{h})
\\
\frac{q}{q-1}\tilde{k}_{\Delta_-}(h)k_{\Delta_+}(\bar{h})& \tilde{k}_{\Delta_-}(h)k_{\Delta_+}(\bar{h})
\end{pmatrix},
\eea
where $h$ and $\bar{h}$ correspond to the different eigenfunctions, respectively.
The coefficient is \cite{Turiaci:2017zwd}:
\bea
\frac{1}{\alpha_{ab}}\equiv(q-\delta_{ab})\alpha^2_{\mathrm{sq}}; \
\alpha_{\mathrm{sq}}\equiv
\frac{(1-\Delta_+)(1-\Delta_-)}
{4\pi^2\tan\big(\frac{\pi}{2}\Delta_+\big)
\cot\big(\frac{\pi}{2}\Delta_-\big)
}.
\eea
The variables are given by:
\bea
k_{\Delta_+}(h)&\equiv&\int_{-\infty}^{\infty} dx_1dx_2\
\frac{\sgn(-x_1)\sgn(1-x_2)\sgn(x_1-x_2)}
{|x_1|^{\Delta_+}|1-x_2|^{\Delta_+}|x_1-x_2|^{(2-\Delta_+-h)}};
\nn\\
k_{\Delta_-}(h)&\equiv&\int_{-\infty}^{\infty} dx_1dx_2\
\frac{1}
{|x_1|^{\Delta_-}|1-x_2|^{\Delta_-}|x_1-x_2|^{(2-\Delta_--h)}}.
\eea
We can solve the spectrum by $\mathrm{det}(I-K)=0$ when the kernel and the identity operator act on the eigenfunctions.

\section{Chaos and Unitarity Bounds}
\label{sec:3}
\noindent
To investigate the theoretical properties in the IR region, we compute the OTOC and the four-point function of four identical fermions.
The exponent of the OTOC has the upper bound given by the $2\pi/\beta$ (chaos bound) for a unitary theory \cite{Maldacena:2015waa}.
We demonstrate that the Lyapunov exponent breaches both the chaos bound and the unitarity bound.

\subsection{Chaos Bound}
\begin{figure}
\centering
\includegraphics[width=1.\linewidth]{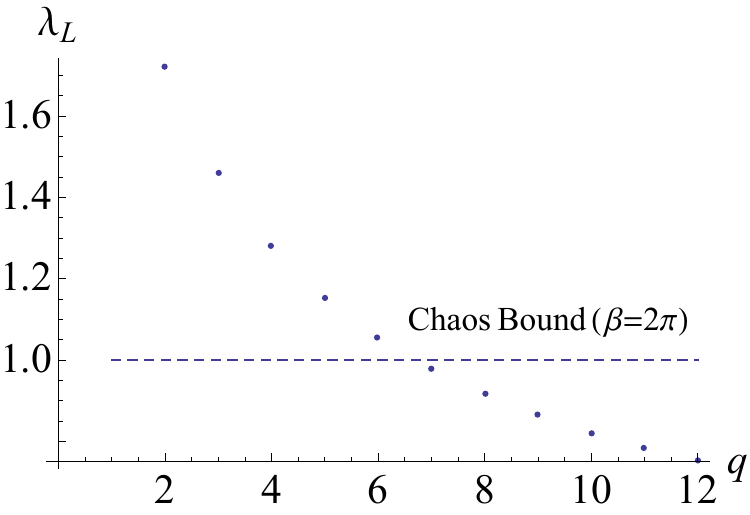}
\caption{The Lyapunov exponents for different $q$.
The chaos bound is violated for $q\le 6$.
The dashed line indicates the chaos bound.
}
\label{Lyapunov}
\end{figure}

\noindent
The four-point function ${\mathcal F}$ can be expanded using a complete basis labeled by the conformal dimensions:
\bea
\label{hls}
h=\frac{1+l}{2}+i s_1; \ \bar{h}=\frac{1-l}{2}+is_1.
\eea
The four-point function concerns the summation of positive integers and the integration range from $-\infty$ to $\infty$,
\bea
{\mathcal F}\sim \sum_{l\in \mathbb{Z}^+}\int_{-\infty}^{\infty} ds_1.
\eea
Following the techniques introduced in Ref. \cite{Murugan:2017eto}, one can rewrite the sum into a contour integral in $\ell$, which (for a fixed $s_1$) picks up the poles at $\mathrm{det}(I-K(h,\bar h))=0$.
The conformal dimensions $h,\bar h$ are related to $\ell, s_1$ as in Eq. \er{hls}.
In this sense, $\ell$ can be determined by $s_1$.
In the large $t$ limit, the integral over $s_1$ concentrates near $s_1=0$.
We obtain the Lyapunov exponent as follows $\lambda_L = l(s_1=0) - 1$ when $\beta = 2\pi$.
The factor $(2\pi)/\beta$ for other finite-temperature values can be derived through dimensional analysis.
Therefore, our analysis remains generalizable.
Our result shows the violation of the chaos bound for $q\le 6$ in Fig. \ref{Lyapunov}.

\subsection{Unitarity Bound}
\begin{figure}
\centering
\includegraphics[width=1.\linewidth]{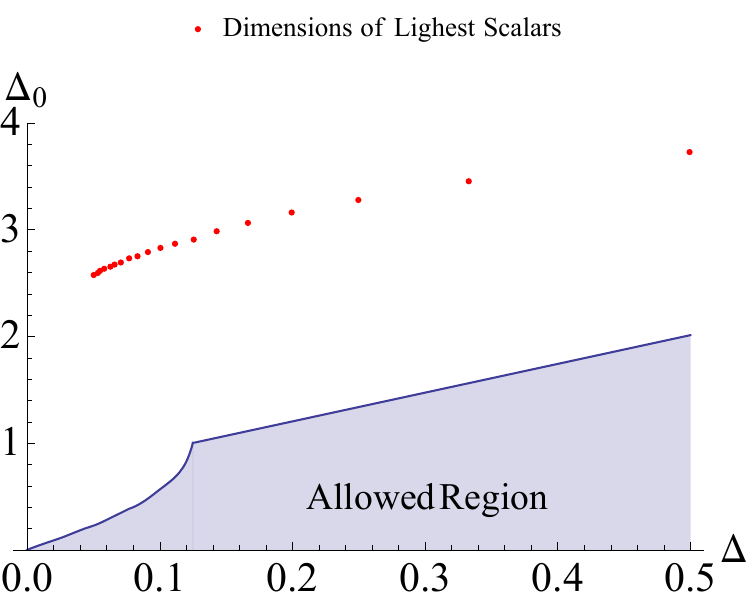}
\caption{The dimensions of the lightest scalars in the spectrum are compared to the shaded allowed region set by the unitarity bound.
}
\label{ub}
\end{figure}

\noindent
We consider the four-point function of four identical fermions
\bea
\langle\psi(x_1)\psi(x_2)\psi(x_3)\psi(x_4)\rangle
=\frac{g(u, v)}{|x_1-x_2|^{2\Delta}|x_3-x_4|^{2\Delta}},
\eea
where
\bea
g(u, v)\equiv \sum_{\mathcal O}(f_{\psi\psi{\mathcal O}})^2g_{\Delta_1, l}(u, v).
\eea
The quantities $g_{\Delta_1, l}$ are referred to as global conformal blocks.
Additionally, the cross ratios are defined as follows:
\bea
u\equiv\frac{|x_1-x_2|^2|x_3-x_4|^2}{|x_1-x_3|^2|x_2-x_4|^2}; \
v\equiv\frac{|x_2-x_3|^2|x_1-x_4|^2}{|x_1-x_3|^2|x_2-x_4|^2}.
\eea
The $\Delta_1$ and $l$ are the dimensions and spins of operators in the $\psi\times\psi$ OPE, respectively.
The explicit forms of $g_{\Delta_1, l}(u, v)$ are expressed in terms of hypergeometric functions \cite{Dolan:2000ut}
\bea
g_{\xD_1,\ell}(z)&= &|z|^{\xD_1-\ell }\times\Big[z^\ell {}_2F_1\left({\frac {\xD_1+\ell} 2},{\frac{\xD_1+ \ell} 2},\xD_1+\ell;z\right)
\nn\\
& &\times {}_2F_1\left({\frac{\xD_1-\ell} 2},{\frac {\xD_1-\ell} 2},\xD_1-\ell;\bar{z}\right) + (z\leftrightarrow \bar{z}) \Big]\,,
\eea
where the cross ratios are expressed in their complex form
\bea
u = z \bar z \,,\quad v = (1-z)(1-\bar z)\,.
\eea
The four-point function exhibits invariance under crossing symmetry, which means: $x_1\leftrightarrow x_3$ or  $x_2\leftrightarrow x_4$.
From this symmetry, we can deduce that
\bea
g(u, v)=\bigg(\frac{u}{v}\bigg)^{\Delta}g(v, u).
\eea
This relationship can be expressed in terms of conformal blocks as follows
\bea
\sum_{\mathcal O}(f_{\psi\psi{\mathcal O}})^2F_{\Delta_1, l}^{\Delta}=0; \
F_{\Delta_1, l}^{\Delta}(u, v)\equiv v^{\Delta}g_{\Delta_1, l}(u, v)-u^{\Delta}g_{\Delta_1, l}(v, u).
\nn\\
\eea
Solving the equation is equivalent to finding the vanishing linearity,
\bea
\sum_{\Delta_1, l}p_{\Delta_1, l}F_{\Delta_1, l}^{\Delta}=0, \ p_{\Delta_1, l}\ge 0.
\eea
The conformal dimension of a scalar field $\Delta_1$ must be smaller than a certain number $\Delta_0$.
We can solve for the smallest scalar conformal dimension $\xD_0$ from the kernel \er{kernel} by requiring $\mathrm{det}(I-K)=0$ ($\ell = h-\bar h = 0$).
Our result shows the violation of the known unitary bound \cite{Rychkov:2009ij} in Fig. \ref{ub}.

\section{Gravitational Dual}
\label{sec:4}
\noindent
We do the field redefinition $\tilde{\Sigma}^a\rightarrow
\tilde{\Sigma}^a
+\epsilon^{\mu\nu}e_{\mu}^a\partial_{\nu}$ 
to separate the reparametrization invariance and non-invariance terms in the effective action
\bea
\frac{S_{\mathrm{eff, 1}}}{N}&=&
-\frac{1}{2}\sum_{a=\pm}\bigg(\ln\det(-\tilde{\Sigma}^a\big)\bigg)
\nn\\
&&
+\frac{1}{2}\sum_{a=\pm}\bigg(\int d^2xd^2\tilde{x}\ \tilde{\Sigma}^a\tilde{G}_a\bigg)
-\frac{J^2}{2q}\int d^2xd^2\tilde{x}\ (\tilde{G}_+)^q(\tilde{G}_-)^q
\nn\\
&&
+\frac{1}{2}\sum_{a=\pm}\int d^2x\ \epsilon^{\mu\nu}e_{\mu}^a\partial_{\nu}\tilde{G}_a
-\epsilon^{\mu\nu}\int d^2x\ e_{\mu}^+e_{\nu}^-.
\eea
The following terms in the effective action break the reparametrization invariance
\bea
\frac{S_{\mathrm{nori}}}{N}=
\frac{1}{2}\sum_{a=\pm}\int d^2x\ \epsilon^{\mu\nu}e_{\mu}^a\partial_{\nu}\tilde{G}_a
-\epsilon^{\mu\nu}\int d^2x\ e_{\mu}^+e_{\nu}^-.
\eea
After we integrate out the $e_{\mu}^{\pm}$, the non-reparametrization terms become that
\bea
\frac{S_{\mathrm{nori, 1}}}{N}=
\frac{1}{4}\epsilon^{\mu\nu}\int d^2x\ \partial_{\mu}\tilde{G}_+\partial_{\nu}\tilde{G}_-.
\eea
\\

\noindent
Our ansatz $G_{\pm, c}$ can be transformed as
\bea
&&
\tilde{G}_{\pm}(x^+, x^-, \tilde{x}^+, \tilde{x}^-)
\nn\\
&=&
\big(f_+^{\prime}(x^+)f_+^{\prime}(\tilde{x}^+)\big)^{\Delta}
\big(f_-^{\prime}(x^-)f_-^{\prime}(\tilde{x}^-)\big)^{\Delta}
\nn\\
&&\times
G_{\pm, c}\big(f_+(x^+), f_-(x^-), f_+(\tilde{x}^+), f_-(\tilde{x}^-)\big)
,
\nn\\
\eea
where $f_{\pm}$ are arbitrary functions.
To obtain the boundary theory as in the 1D fermionic SYK model, we do the expansion for $\tilde{G}_{\pm}(x^{\pm}, \tilde{x}^{\pm})$ with $x^{\pm}-\tilde{x}^{\pm}\ll 1$ \cite{Maldacena:2016hyu}.
The low-energy theory is given by
\bea
&&
\frac{1}{2}\epsilon^{\mu\nu}\int d^2xd^2\tilde{x}\ \partial_{\mu}\tilde{G}_{+, s}(x^{\pm}, \tilde{x}^{\pm})\partial_{\nu}\tilde{G}_{-,s}(x^{\pm}, \tilde{x}^{\pm})
\delta(x^+-\tilde{x}^+)\delta(x^--\tilde{x}^-)
\nn\\
&=&
\frac{1}{4}\int d^2x^{\pm}d^2\tilde{x}^{\pm}\ \big(\partial_{+}\tilde{G}_{+, s}(x^{\pm}, \tilde{x}^{\pm})\partial_{-}\tilde{G}_{-, s}(x^{\pm}, \tilde{x}^{\pm})
\nn\\
&&
-\partial_{-}\tilde{G}_{+, s}(x^{\pm}, \tilde{x}^{\pm})\partial_{+}\tilde{G}_{-, s}(x^{\pm}, \tilde{x}^{\pm})\big)
\nn\\
&&\times
\delta(x^+-\tilde{x}^+)\delta(x^--\tilde{x}^-),
\eea
where $\tilde{G}_{\pm, s}\equiv\tilde{G}_{\pm}-\tilde{G}_{\pm, c}$.
Thus, we derive the double Schwarzian theory from the most prominent singular term
\bea
&&
\frac{1}{4}\int d^2x^{\pm}d^2\tilde{x}^{\pm}\ \big(\partial_{+}\tilde{G}_{+, s}(x^{\pm}, \tilde{x}^{\pm})\partial_{-}\tilde{G}_{-, s}(x^{\pm}, \tilde{x}^{\pm})
\nn\\
&&
-\partial_{-}\tilde{G}_{+, s}(x^{\pm}, \tilde{x}^{\pm})\partial_{+}\tilde{G}_{-, s}(x^{\pm}, \tilde{x}^{\pm})\big)
\nn\\
&&\times
\delta(x^+-\tilde{x}^+)\delta(x^--\tilde{x}^-)\propto \int d^2\tilde{x}^{\pm}\ \{f_+, \tilde{x}^+\}\{f_-, \tilde{x}^-\}.
\eea
The integration $\int d^2x^{\pm}$ contributes to the coefficient of the double Schwarzian theory.
The double Schwarzian theory is the boundary theory of the AdS$_3$ Einstein gravity with a finite radial cut-off at the classical level \cite{Freidel:2008sh,McGough:2016lol}.

\section{Higher Dimensions}
\label{sec:5}
\noindent
To explore the higher-dimensional AdS/CFT correspondence, we provide the generalization to (Euclidean) $d$-dimensions
\bea
S_{\mathrm{UV}}&=&\sum_{j_1=1}^N\int d^dx\ \epsilon^{\mu_1\mu_2\cdots\mu_d}
(\psi_1^{j_1}\partial_{\mu_1}\psi_1^{j_1})
(\psi_2^{j_1}\partial_{\mu_2}\psi_2^{j_1})
\cdots
(\psi_d^{j_1}\partial_{\mu_d}\psi_d^{j_1});
\nn\\
S_{\mathrm{IR}}&=&\sum_{1\le i_<  i_2<\cdots<i_q<j_1<j_2<\cdots<j_q\le N}\int d^dx \ J_{i_1,i_2, \cdots, i_q, j_1, j_2, \cdots, j_q, \cdots}
\nn\\
&&
\times
(\psi_1^{i_1}\psi_1^{i_2})\cdots (\psi_1^{i_{q-1}}\psi_1^{i_q})(\psi_2^{j_1}\psi_2^{j_2})\cdots(\psi_2^{i_{q-1}}\psi_2^{j_q})\cdots,
\eea
where $\psi_{1}, \psi_2, \cdots, \psi_d$ are $m$-component of $d$-dimensional Majorana fermions.
We only concern  $d\in 4\mathbb{Z}^+$ or sympletic Majorana Weyl spinor ($d=4, 12, 20, \cdots$) and Majorana Weyl spinor ($d=8, 16, 24, \cdots$).
When $d=4$, the fermion fields are two-component symplectic Majorana spinors ($m=2$).
For $d=8$, the fermion fields are eight-component Majorana Weyl spinors ($m=8$).
$J_{i_1,i_2, \cdots, i_q, j_1, j_2, \cdots, j_q\cdots}$ also follows a Gaussian random distribution
\bea
\langle J_{i_1,i_2, \cdots, i_q, j_1, j_2, \cdots, j_q, \cdots}J_{i_1,i_2, \cdots, i_q, j_1, j_2, \cdots, j_q, \cdots}\rangle=\frac{J^2(q-1)!q!}{N^{2q-1}}.
\eea
This model has higher derivatives in the kinematic term $S_{\mathrm{UV}}$, resulting in dimensionless fermion fields.
Thus, the fixed coupling constant also has a positive energy dimension $[J]=d$, which ensures renormalizability.
The interaction term also dominates at the IR limit.
When considering the IR limit, the dominant term exhibits reparametrization and local Lorentz symmetries.
The large-$N$ SD equation has a form similar to that of the lower-dimensional case.
We illustrate the large-$N$ SD equation for $d=4$ up to the derivative corrections.
Now, our fields with the spinor indices ($a_1, a_2$) can be rewritten as that
\bea
(\tilde{{\mathcal G}}_a)_{a_1a_2}\equiv\epsilon_{a_1a_2}\tilde{G}_a; \ (\tilde{{\mathcal S}}_a)_{a_1a_2}\equiv\epsilon_{a_1a_2}\tilde{\Sigma}_a.
\eea
The large-$N$ SD equation becomes:
\bea
\bar{G}_{a}\bar{\Sigma}_{a}\approx -1; \
\tilde{\Sigma}^{a}(x)=J^2\tilde{G}_{a}^{q-1}(x)\tilde{G}_{b_1}^q(x)\cdots \tilde{G}_{b_{d-1}}^q(x).
\eea
We can still use the conformal solution as the ansatz:
\bea
\tilde{G}_a=b\frac{1}{|x^{\mu}x_{\mu}|^{\Delta}}; \
\tilde{\Sigma}_a=J^2 b^{qd-1}\frac{1}{|x^{\mu}x_{\mu}|^{(d-\Delta)}}.
\eea
We can use $\tilde{\Sigma}^{a}(x)=J^2\tilde{G}_{a}^{q-1}(x)\tilde{G}_{b_1}^q(x)\cdots \tilde{G}_{b_{d-1}}^q(x)$
to show $\Delta=1/q$.
The coefficient $b$ can be determined by $\bar{G}_{a}\bar{\Sigma}_{a}\approx -1$.
\\

\noindent
Our generalization guarantees the UV completion.
Therefore, we can reach our goal of studying the emergence of the graviton from a UV-complete QFT, not restricted to CFTs.
Our construction in higher dimensions is computable, just as in lower dimensions.
Computing the Lyapunov exponent to investigate the unitary issue, similar to the 2D case, is expected to be an interesting approach.

\section*{Acknowledgments}
\noindent 
XH acknowledges the NSFC Grants (Grants No. 12247103 and No. 12475072). 
CTM thanks Nan-Peng Ma for his encouragement. 
The authors would like to thank the Isaac Newton Institute for Mathematical Sciences, Cambridge, for support and hospitality during the programme Quantum field theory with boundaries, impurities, and defects, where work on this paper was undertaken. 
This work was supported by EPSRC grant EP/Z000580/1.


\end{document}